\documentclass[twocolumn,aps,prl,showpacs,amsmath,superscriptaddress,longbibliography,notitlepage]{revtex4-2}
\usepackage{amssymb}
\usepackage{mathrsfs}
\usepackage{graphicx}
\usepackage{float}
\usepackage[caption=false]{subfig}
\usepackage{epstopdf}

\usepackage[normalem]{ulem}
\usepackage{verbatim}
\usepackage{xcolor}
\usepackage{bm}

\begin{document}

\title{Reconfigurable Defect States in Non-Hermitian Topolectrical Chains with Gain and
Loss}
\author{S. M. Rafi-Ul-Islam}
\email{rafiul.islam@u.nus.edu}
\affiliation{Department of Electrical and Computer Engineering, National University of Singapore, Singapore 117583, Republic of Singapore}
\author{Zhuo Bin Siu}
\email{elesiuz@nus.edu.sg}
\affiliation{Department of Electrical and Computer Engineering, National University of Singapore, Singapore 117583, Republic of Singapore}
\author{Md. Saddam Hossain Razo}
\email{shrazo@u.nus.edu}
\affiliation{Department of Electrical and Computer Engineering, National University of Singapore, Singapore 117583, Republic of Singapore}
\author{Mansoor B.A. Jalil}
\email{elembaj@nus.edu.sg}
\affiliation{Department of Electrical and Computer Engineering, National University of Singapore, Singapore 117583, Republic of Singapore}

\begin{abstract}
We investigate the interplay between the non-Hermitian skin effect (NHSE), parity-time (PT) symmetry, and topological defect states in a finite non-Hermitian Su-Schrieffer-Heeger (SSH) chain. In the conventional NHSE regime, non-reciprocal hopping leads to an asymmetric localization of all eigenstates at one edge of the system, including the bulk and topological edge states. However, the introduction of staggered gain and loss restores the symmetric localization of topological edge states while preserving the bulk NHSE. We further examine the response of defect states in this system, demonstrating that their spatial localization is dynamically controlled by the combined effects of NHSE and PT symmetry. Specifically, we identify three distinct regimes in which the defect states localize at the defect site, shift to the system's edges, or become completely delocalized. These findings extend beyond previous works that primarily explored the activation and suppression of defect states through gain-loss engineering. To validate our theoretical predictions, we propose an experimental realization using a topolectrical circuit, where non-Hermitian parameters are implemented via impedance converter-based non-reciprocal elements. Circuit simulations confirm the emergence and tunability of defect states through voltage and admittance measurements, providing a feasible platform for experimental studies of non-Hermitian defect engineering. Our results establish a route for designing reconfigurable non-Hermitian systems with controllable topological defect states, with potential applications in robust signal processing and sensing.
\end{abstract}

\maketitle
\section{Introduction} 
In recent years, non-Hermitian systems \cite{leykam2017edge,rafi2024saturation,ashida2020non,siu2023terminal,rafi2024exceptional,bergholtz2021exceptional,rafi2023twisted,gong2018topological} have attracted significant attention owing to their unique properties and potential applications in various fields such as photonics \cite{yan2023advances,feng2017non,longhi2018parity}, condensed matter physics \cite{santra2002non,wang2018valley,bagarello2016non,rafi2024knots}, and quantum computing \cite{zhang2019time,longhi2010optical,bender2007faster,ju2019non}. These systems have complex eigenvalues and non-orthogonal eigenvectors, making them fundamentally different from their Hermitian counterparts \cite{hahn2016observation,el2018non,sahin2024protected,zhang2023anomalous}.

One intriguing phenomenon in non-Hermitian systems is the non-Hermitian skin effect (NHSE) \cite{okuma2020topological,rafi2024twisted,siu2023terminal,rafi2024dynamic,rafi2025critical,li2020critical,zhang2022universal,song2019non}, which refers to the localization of bulk states near the boundaries of the system. The NHSE has been observed in various non-Hermitian systems such as non-Hermitian photonic lattices \cite{zhong2021nontrivial,zhu2020photonic,song2020two}, topolectrical (TE) circuits \cite{rafi2020topoelectrical,liu2021non,rafi2020realization,sahin2023impedance,rafi2022interfacial,helbig2020generalized,rafi2023valley,rafi2021non,hofmann2020reciprocal,rafi2022unconventional,rafi2021topological,xu2021coexistence}, non-Hermitian topological insulators \cite{zhang2021observation,rafi2023conductance,kawabata2020higher,rafi2023engineering,lin2021square}, and non-Hermitian quantum systems \cite{song2019non,rafi2024anomalous,okuma2021quantum}.

Topological edge states are a hallmark of topological phases of matter \cite{rafi2022system,fujita2011gauge,rafi2024chiral,obana2019topological,rafi2023valley,hafezi2013imaging,rafi2022type}, and have been extensively studied in Hermitian systems. In recent years, the study of topological edge states has been extended to non-Hermitian systems \cite{yuce2018edge,esaki2011edge}. The coexistence of topological edge states and NHSE in non-Hermitian systems has been shown to lead to novel phenomena, such as the selective enhancement of topological zero modes \cite{poli2015selective}. Another intriguing feature of non-Hermitian systems is the existence of topological defect states \cite{blanco2016topological,lang_effects_2018,poli2015selective,cui2020localized,kong2020energy,garmon2021reservoir,teo2017topological,stegmaier2021topological,barkeshli2013classification}, which are localized states that appear at the interface of two regions with different topological properties. These states have been studied in various non-Hermitian systems, such as non-Hermitian photonic lattices \cite{pan2018photonic} and non-Hermitian topological insulators \cite{stegmaier2021topological}.  Recently, Stegmaier \textit{et al.} (Ref. \onlinecite{stegmaier2021topological}) demonstrated the critical role of system symmetry in the emergence of defect states within a configuration comprising two non-Hermitian chains, each characterized by staggered gain/loss terms and connected via a central defect site. Specifically, they revealed that defect states are present when the system exhibits parity-time (PT) and anti-PT (APT) symmetries, but are absent under conditions of broken-PT (BPT) symmetry. However, the model employed by Stegmaier et al. did not incorporate any coupling asymmetry within the chain. Consequently, the interplay between defect states, topological edge states and the localization effects attributed to NHSE, remained unexplored.

In this paper, we investigate the effect of the NHSE on the localization properties of defect and edge states in non-Hermitian systems. We show that the NHSE drags not only the bulk states but also the edge states towards the boundary, resulting in the localization of both types of eigenstates near the boundary.  We further explore the possibility of enhancing the localization properties of the edge states by introducing staggered gain and loss potential to the sublattice sites. This results in the superimposition of the NHSE and edge state localization  at opposite edges.

We further analyze the effect of the gain/loss term on  defect states. We introduce balanced gain/loss terms on the A/B sublattice sites of a Su-Schrieffer-Heeger (SSH) chain while leaving the defect states passive. By modulating the gain/loss terms, we observe different types of eigenstates and localization properties depending on the PT symmetry of the system. Interestingly, we find that defect states appear only in the PT- and APT-symmetric cases and are suppressed   in the broken PT-symmetric case. We explain analytically the observed correlation between the emergence and localization of the defect states and the PT symmetry. Additionally, we propose an experimental realization of our model using a non-hermitian TE circuit. This circuit design validates our theoretical predictions through its electrical characteristics, such as voltage and admittance responses, by demonstrating the presence and control of defect states via non-Hermitian parameters. The capability to switch topological defect states and adjust their localization has significant implications for defect engineering in non-Hermitian systems.

\section{Characteristics of pristine non-Hermitian SSH chain with loss and gain}
\begin{figure}[ht!]
  \centering
\includegraphics[width=0.47\textwidth]{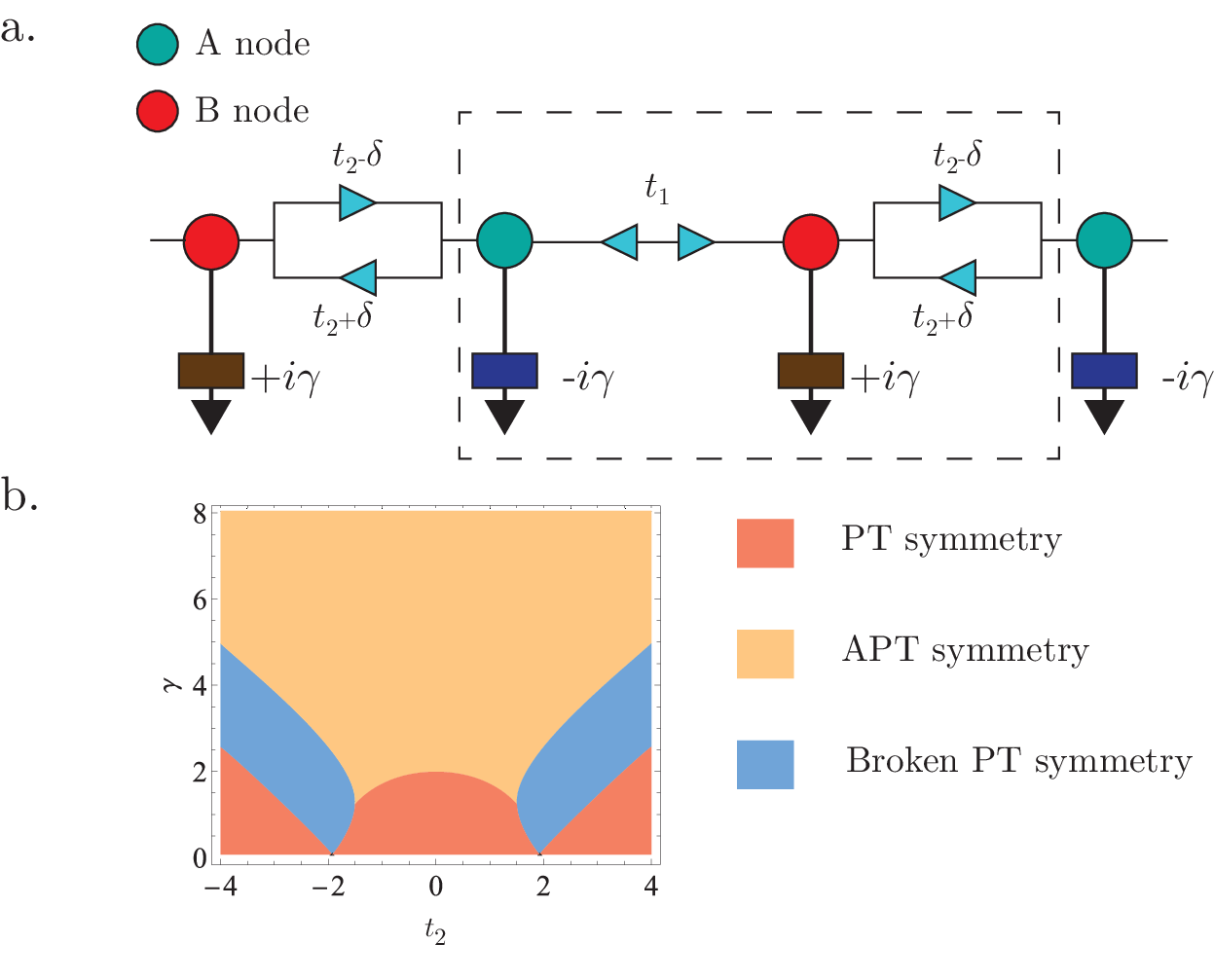}
  \caption{Schematic and phase diagram of a non-Hermitian system with staggered gain and loss terms. (a) The lattice consists of two sublattice nodes A and B with staggered gain and loss terms ($+i\gamma$ and $-i\gamma$), respectively. The dotted box demarcates a single unit cell. The intra-unit cell coupling $t_1$ is reciprocal while the intercell coupling $t_2 \pm \delta$ is non-reciprocal. (b) Phase diagram as a function of inter-cell coupling $t_2$ and gain/loss factor $\gamma$  depicting  the three different symmetry cases of PT-symmetric (only real eigenenergies), broken PT-symmetric (mixture of real and imaginary eigenenergies), and APT-symmetric (only imaginary eigenenergies). Common parameters: $t_1=1.2$, $\delta=-1.5$. }
  \label{fig1}
\end{figure}

We investigate the emergence of the NHSE and topological edge states in a generic one-dimensional lattice model with two sublattice nodes denoted as A and B nodes (refer to Fig. \ref{fig1}a). Specifically, we focus on a modified non-Hermitian SSH chain with reciprocal intra-chain ($t_1$) and non-reciprocal inter-chain ($t_2 \pm \delta$) coupling in which gain ($-i \gamma$)  and loss ($i \gamma$) are present at the A and B nodes, respectively. Under periodic conditions, the Bloch Hamiltonian $H(k)$ of the system is given by
\begin{widetext}
\begin{equation}
H(k)= (t_1+t_2 \cos k_x - i \delta \sin k_x) \sigma_x +(t_2 \sin k_x  + i \delta \cos k_x) \sigma_y + i \gamma \sigma_z,
\label{Eq1Ham}
\end{equation}
\end{widetext}
where $\sigma_i$ is the $i$th Pauli matrix, $t_1$ and $t_2$ denote the intra- and inter-cell couplings, respectively, and $\delta$ the degree of non-reciprocity in the inter-unit cell couplings, which may be achieved, for example, via the use of negative impedance converters at current inversion (INICs) in a TE realization \cite{rafi2021topological,rafi2022critical,rafi2022unconventional,rafi2022interfacial}. As above-mentioned, $\gamma$ represents the staggered gain and loss terms. The corresponding  eigenvalues are given by
\begin{widetext}
\begin{equation}
E=\pm \sqrt{t_1^2+t_2^2-\gamma^2-\delta^2+ 2t_1 t_2 \cos k_x - 2 i t_1 \delta \sin k_x}.
\end{equation}
\label{eq2Adm}
\end{widetext}

Unless otherwise stated, we assume the case where $|t_1| < |\sqrt{t_2^2-\delta^2}|$ and that $|\delta|<|t_2|$.  All the eigenenergies of the bulk states in a finite open chain are real when $\gamma<||t_1|-\sqrt{t_2^2-\delta^2}|$, which corresponds to the PT-symmetric case. However, at large values of the gain/loss term, i.e., $\gamma > ||t_1|+\sqrt{t_2^2-\delta^2}|$, the eigenenergies become imaginary in the anti PT-symmetric phase. At intermediate values of $\gamma$ at which $||t_1|-\sqrt{t_2^2-\delta^2}|<\gamma<||t_1|+\sqrt{t_2^2-\delta^2}|$, the eigenvalue spectrum contains a mixture of real and imaginary eigenenergies, which is referred to as the broken PT-symmetric case. These three phases are delineated in the  phase diagram drawn in Fig. \ref{fig1}b.
\begin{figure*}[ht!]
  \centering
\includegraphics[width=0.9\textwidth]{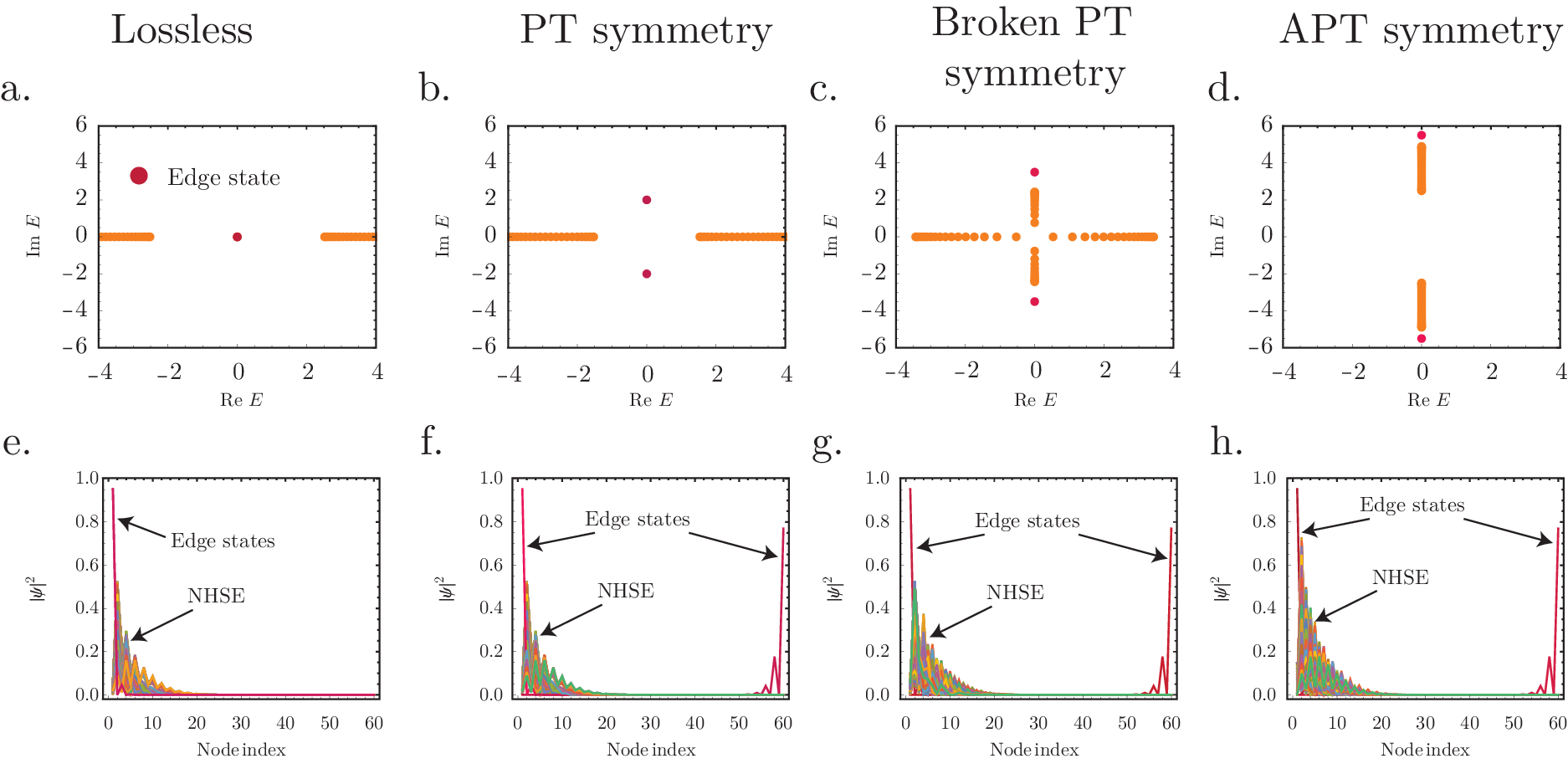}
  \caption{Complex energy and eigenstate distributions of the non-Hermitian SSH model with asymmetric couplings and staggered gain and loss terms corresponding to different classes of symmetry. (a,e) In the absence of gain/loss ($\gamma=0$),  the energy eigenvalues are real (a) and both topological edge states are localized at the preferred NHSE accumulation edge (e). (b,f) The PT-symmetric case ($\gamma =2$), where the energy eigenvalues are still real (b) and the edge states are symmetrically localized at the two ends of the chain (f). (c,g) The broken PT-symmetric case ($\gamma= 3.5$), where (c) there is now a mixture of real and imaginary energy eigenvalues and (g) the two topological edge states are localized at opposite edges. (d,h) The right panel (d) displays the APT-symmetric case ($\gamma=5.5$) where the energy eigenvalues are purely imaginary and (h) the edge states  are localized at both edges similar to (g). The parameters used are $t_1=1.2$, $t_2=4$, and $\delta=-1.5$.  }
  \label{fig2}
\end{figure*} 
 
We analyze the evolution of the edge and NHSE states for the three different PT  symmetry cases by plotting the complex eigenenergy  and eigenstate spatial distributions under varying gain/loss strengths $\gamma$ in Fig. \ref{fig2}.

We first consider the case with $\gamma=0$. The Hamiltonian in Eq. \eqref{Eq1Ham} exhibits a chiral symmetry, i.e., $\mathcal{C}H(k)|_{\gamma=0} \mathcal{C}^{-1}=-H(k)$ with $\mathcal{C}=\sigma_z$. The chiral symmetry protects the topologically nontrivial states, which are characterized by a winding number. One way to calculate this winding number is to convert Eq. \eqref{Eq1Ham} into a Hermitian form via a similarity transform as described below.  

In the absence of the loss/gain term (i.e. $\gamma = 0$), the Hamiltonian in Eq. \eqref{Eq1Ham} under OBC  can be written in the real space basis as
\begin{align}
	H_0 = \sum_m & |m\rangle| t_1 \sigma_x \langle m| \nonumber \\
& + |m\rangle |\mathrm{A}\rangle (t_2 + \delta) \langle\mathrm{B}|\langle m-1| \nonumber \\
& + |m\rangle |\mathrm{B}\rangle (t_2 - \delta) \langle\mathrm{A}|\langle m+1| \label{eqH0} 
\end{align} 
where $|m\rangle$ is the basis state for the $m$th unit cell, $|\mathrm{A} / \mathrm{B}\rangle$ is the basis state  for the A / Bth sublattice, and $\sigma_x = |\mathrm{A}\rangle\langle\mathrm{B}| +  |\mathrm{B}\rangle\langle\mathrm{A}|$. $H_0$ can be converted into a Hermitian Hamiltonian $\overline{H}_0$ via a similarity transformation operator $S$:
\begin{equation} 
	\overline{H}_0 = S^{-1}H_0S \label{SH0S}
\end{equation}
where
\begin{align}
	\overline{H}_0 = \sum_m & |\overline{m}\rangle t_1 \sigma_x \langle \overline{m}| \nonumber \\
& + |\overline{m}\rangle |\mathrm{A}\rangle \overline{t}_2  \langle\mathrm{B}|\langle\overline{ m-1}| \nonumber \\
& + |\overline{m}\rangle |\mathrm{B}\rangle \overline{t}_2  \langle\mathrm{A}|\langle\overline{ m+1}| \label{eqHb0}
\end{align}
and
\begin{equation}
   S = \sum_{m} |m\rangle \left( \frac{t_2+\delta}{t_2-\delta} \right)^{m/2} \mathbf{I}_2 \langle\overline{ m}| \label{Smat} 
\end{equation}
\begin{equation}
    S^{-1} = \sum_{m} |\overline{m}\rangle \left( \frac{t_2-\delta}{t_2+\delta}\right)^{m/2} \mathbf{I}_2 \langle m|.
\end{equation}
It can be seen by a direct comparison of Eqs. \eqref{eqH0} and \eqref{eqHb0} that $\overline{t}_2$ in the latter corresponds to  $\sqrt{ (t_2+\delta)(t_2-\delta)}$ in the former (in other words, the interchain coupling of $t_2$ is replaced by the  $\sqrt{t_2^2-\delta^2}$. In the original SSH chain with gain/loss but without coupling asymmetry (i.e., $ \delta=0$), we can obtain the completely real or imaginary eigenspectrum if $\gamma\ <\left|t_1-t_2\right|$  and  $\gamma\ >\left|t_1+t_2\right|$, respectively.  Hence, for the non-Hermitian SSH chain if we replace the $ t_2$ by the $ \sqrt{t_2^2-\delta^2}$, the condition for entirely real and complex eigenspectrum is then given by  $\gamma<\left|\left|t_1\right|-\sqrt{t_2^2-\delta^2}\right|$ and $ \gamma>\left|\left|t_1\right|+\sqrt{t_2^2-\delta^2}\right|$, respectively). Eqs. \eqref{eqH0} and \eqref{eqHb0} share the same eigenvalues while their corresponding eigenvectors are scaled relative to each other by an exponential factor of $((t_2+\delta)(t_2-\delta))^{m/2}$. A consequence of this is that topological phase of the non-Hermitian Hamiltonian in Eq. \eqref{eqH0} can be predicted by calculating the winding number of its Hermitian counterpart in Eq. \eqref{eqHb0}. This winding number is given by
\begin{equation}
\begin{split}
\mathcal{W}& =\frac{1}{2\pi} \int_{-\pi}^{\pi} \frac{\mathrm{d}}{\mathrm{d}k_x} \tan^{-1}\left(\frac{\overline{t}_2 \sin k_x }{t_1+\overline{t}_2 \cos k_x}\right) \\
& =\begin{cases}
    1 & \text{if $|t_1| < |\sqrt{t_2^2-\delta^2}|$} \\    
    0 & \text{otherwise}
  \end{cases}.
\label{eq3Winding}
\end{split}
\end{equation}

For the case where the winding number is nonzero, two topological zero-energy edge states appear in the complex energy spectrum (see Fig. \ref{fig2}a where the two degenerate edge states occur at $E = 0$.). The asymmetry in the inter-chain couplings results in the exponential localization of all the eigenstates, which is a direct consequence of NHSE (see Fig. \ref{fig2}e-h).  For a bulk state, this exponential accumulation arises from the shift of wavevector from a real value, $k\in \mathcal{R}$ under PBC to a complex value of $k+i\kappa$ under OBC, where $\kappa$ is the inverse decay length related to the non-Bloch factor, which is given by $\beta=\lambda e^{ik}$, $\lambda =\exp(-\mathrm{Im} k)$.

In Fig. \ref{fig2}, the given system parameters yield a $\lambda$ value in the range $|\lambda|<1$, which results in all bulk states to be localized at the left edge (Figs. \ref{fig2}e-h). It is worth noting that the NHSE causes both of the topological edge states to be displaced  towards the preferred edge (i.e., the left edge in the case depicted by Fig. \ref{fig2}e) when $\gamma=0$. This behavior is different from the Hermitian case where the two edge states are localized at two opposite ends \cite{yao2018edge}. In contrast, a non-zero $\gamma$ restores the localization of the edge states to the two end nodes (Figs. \ref{fig2}f-h) seen in the Hermitian case. The restoration is independent of the NHSE localization of the bulk states  remains largely unchanged by the insertion of the gain/loss term. Additionally, the alternating gain/loss term does not significantly affect the  decay lengths of the bulk or topological edge states (Fig. \ref{fig2}f-h). Therefore, one can surmise that it is the presence of   onsite gain/loss potential $\gamma$ rather than its sign which causes one of the topological edge states to be spatially separated from the NHSE localized bulk  states.

We will now analytically explain the observed behavior of the topological edge states by considering  the concept of the edge Hamiltonian. For a finite system of $N$ unit cells described by the Hermitian Hamiltonian $\overline{H}$, the edge Hamiltonian $\overline{H}_{\mathrm{edge}}$ is constructed by projecting $\overline{H}$ into the basis of $|\overline{b_\pm}\rangle$,which correspond to the  normalized wavefunctions at the first $N$ unit cells of the edge states for semi-infinite systems localized along the left and right edges truncated to $N$ unit cells.  Explicitly, the $|\overline{b_\pm}\rangle$ basis states are given by 
\begin{align}
	|\overline{b_+}\rangle &= \mathcal{N}_+\sum_{m=0}^{N-1} \left( -\frac{t_2}{t_1}\right)^m |\mathrm{B}\rangle |\overline{m}\rangle \label{bp} \\
	|\overline{b_-}\rangle &= \mathcal{N}_-\sum_{m=0}^{N-1} \left( -\frac{t_1}{t_2}\right)^m |\mathrm{A}\rangle |\overline{m}\rangle \label{bm}
\end{align}
where 
\begin{equation}
	\mathcal{N}_\pm = \sqrt{ \frac{1 – \left(\frac{t_2}{t_1}\right)^{\pm 2}}{1-\left(\frac{t_2}{t_1}\right)^{\pm 2N}}} \label{Npm}
\end{equation}
are normalization factors introduced so that $\langle \overline{b_{\pm}}|\overline{b_{\pm}}\rangle = 1$, and the $N$ unit cells are numbered from 0 to $N-1$. When $\gamma=0$, $\overline{H}_{\mathrm{edge}}$ is then given by 
\begin{equation}
	\overline{H}_{\mathrm{edge}} \equiv \begin{pmatrix} \langle \overline{b_+}|\overline{H}|\overline{b_+}\rangle & \langle\overline{ b_+}|\overline{H}|\overline{b_-}\rangle \\ \langle \overline{ b_-}|\overline{H}|\overline{b_+}\rangle & \langle\overline{ b_-}|\overline{H}|\overline{b_-}\rangle \end{pmatrix} = \mathcal{N}_+\mathcal{N}_- t_1\sigma_x. \label{eqHbarEdge}
\end{equation}

The eigenvalues of $\overline{H}_{\mathrm{edge}}$, $\pm \mathcal{N}_+\mathcal{N}_-t_1$, are then approximately the eigenenergies of the topological edge states of Eq. \eqref{eqHb0} while its eigenvectors $|\overline{\pm x}\rangle = \frac{1}{\sqrt{2}} (|\overline{b_+}\rangle \pm |\overline{b_-}\rangle)$ would approximate  the corresponding wavefunctions of the edge states. Note that the similarity transformation of Eq. \eqref{SH0S} implies that if $|\overline{\pm x}\rangle$ is an eigenstate of $\overline{H}_0$, then $|\pm x\rangle \equiv S|\overline{\pm x}\rangle$ is an eigenstate of $H_0$. By using the fact that $|\pm x\rangle = S|\overline{\pm x}\rangle = \frac{1}{\sqrt{2}} (S|\overline{b_+}\rangle \pm |S\overline{b_-}\rangle$ where $S$ is defined in Eq. \eqref{Smat} and $|\overline{b_\pm}\rangle$ are defined in Eqs. \eqref{bp} and \eqref{bm}, respectively, $|\pm x\rangle$ is then given  by 
\begin{equation}
	|\pm x\rangle = \frac{1}{\sqrt{2}} \sum_{m=0}^{N-1}|m\rangle \begin{pmatrix}  \left(  -\frac{t_2(t_2 + \delta)}{t_1(t_2-\delta)} \right)^m  \\ \pm  \left(  -\frac{t_1(t_2 + \delta)}{t_2(t_2-\delta)} \right)^m \end{pmatrix}.
\end{equation}    

Depending on the signs of $\mathrm{Ln}|\left(\frac{t_2(t_2 + \delta)}{t_1(t_2-\delta)}\right)|$ and $\mathrm{Ln}|\left(\frac{t_1(t_2 + \delta)}{t_2(t_2-\delta)}\right)|$, the corresponding topological edge state may be localized at the left (right) edge for positive (negative) values of the logarithm. For the parameter set in Fig. \ref{fig2}e,  both of the topological edge states are localized at the left edge.

We now include the finite gain / loss term, i.e., the Hamiltonian is modified  as 
\begin{equation}
	H = H_0 + \sum_m |m\rangle i\gamma\sigma_z \langle m|.
\end{equation}
Projecting the above into the basis of $|\pm x\rangle$ , we obtain
\begin{align}
	&\tilde{H}_{\mathrm{edge}}  \nonumber \\
\equiv& \begin{pmatrix} \langle +x|H|+x\rangle & \langle +x|H|-x\rangle  \\ \langle -x|H|+x\rangle & \langle -x|H|-x\rangle \end{pmatrix} \\
=& \mathcal{N}_+\mathcal{N}_- t_1\sigma_z + i\gamma\sigma_x. \label{tildeHedge}
\end{align}
It can be observed from the definitions of $\mathcal{N}_\pm$ in Eq. \eqref{Npm} that as $N \rightarrow \infty$, $\mathcal{N}_+ \rightarrow 0$ and $\mathcal{N}_- \rightarrow \sqrt{1-(t_2/t_1)^2}$. The ratio of the magnitude of the $\mathcal{N}_+\mathcal{N}_- t_1\sigma_z$ term relative to that of the $i\gamma\sigma_x$ term therefore becomes smaller as the system size increases, and the eigenvalues of Eq. \eqref{tildeHedge} $\pm \sqrt{(\mathcal{N}_+\mathcal{N}_- t_1)^2-\gamma^2}$, which correspond to the eigenenergies of the edge states, switch from real to imaginary and approach $\pm i\gamma$. This trend can be observed in Fig. \ref{fig2}b to Fig. \ref{fig2}d. Moreover, in the limit of large $N$, the eigenvectors of Eq. \eqref{tildeHedge} approach $\frac{1}{\sqrt{2}} (|+x\rangle \pm |-x \rangle = |b_\pm \rangle$ where
$|b_\pm \rangle \equiv S|\overline{b}\pm \rangle$ and take the explicit forms of 
\begin{align}
	|b_+\rangle &= \mathcal{N}_+ \sum_{m=0}^{N-1} |m\rangle|\mathrm{B}\rangle \left( -\frac{t_2}{t_1}\sqrt{\frac{t_2+\delta}{t_2-\delta}} \right)^m,  \\
	|b_-\rangle &= \mathcal{N}_- \sum_{m=0}^{N-1} |m\rangle|\mathrm{A}\rangle \left( -\frac{t_1}{t_2}\sqrt{\frac{t_2+\delta}{t_2-\delta}} \right)^m.
\end{align}
Under our assumptions of $|\delta|<|t_2|$ and $|t_1| < |\sqrt{t_2^2-\delta^2}|$, the 	factor of $\frac{t_2}{t_1}\sqrt{\frac{t_2+\delta}{t_2-\delta}}$ in $|b_+\rangle$ has a magnitude that is larger than 1, which implies that $|b_+\rangle$ with an eigenenergy near $i\gamma$ will be localized along the right edge. If $|\frac{t_1}{t_2}| < \sqrt{\frac{t_2-\delta}{t_2+\delta}}$, then $|b_-\rangle$, which has an eigenenergy of approximately $-i\gamma$,will be localized along the left edge opposite from $|b_+\rangle$. Thus, under this condition, the introduction of a finite gain / loss term will result in the recovery of the conventional topological edge state configuration, i.e., with the two edge states  localized at opposite edges of  a sufficiently large system, as seen in Figs. \ref{fig2}f – g. 


\begin{figure}[ht!]
  \centering
    \includegraphics[width=0.48\textwidth]{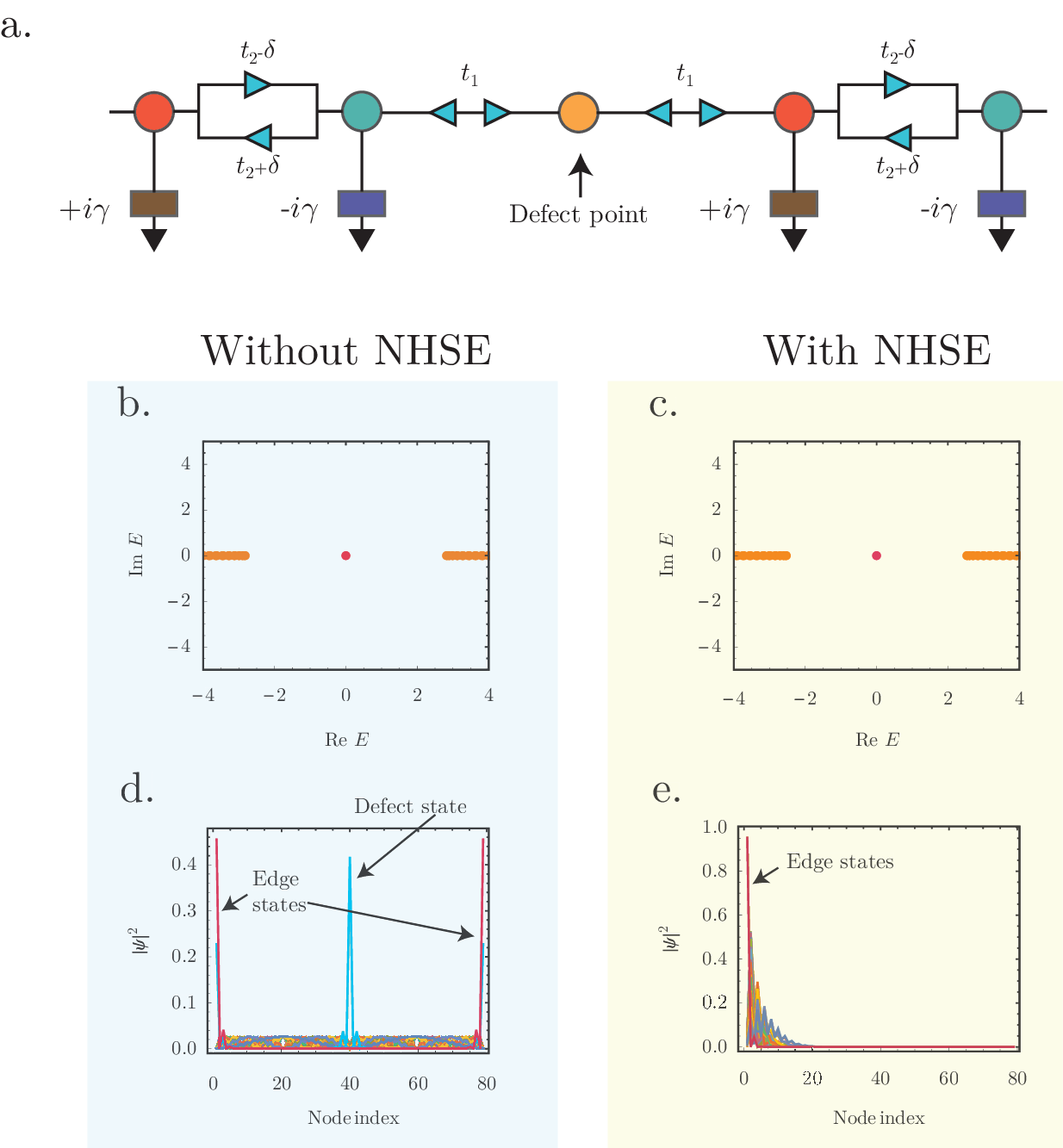}
  \caption{(a) Schematic of the defective non-Hermitian chain. The SSH chain contains a defective node (marked in orange)  that disrupts the couplings between its neighboring nodes. (b,d) Panel (b) shows the complex eigenvalue distribution with zero-energy edge and defect states in the absence of NHSE at $\delta=0$ and panel (d) shows the corresponding eigenstate distribution with topological defect states localized at the defect point and two oppositely localized edge states. (c, e) Panel (c) shows that the eigenspectrum remains almost the same when NHSE is introduced by setting $\delta$ to a finite value of $\delta=-1.5$. (e) The corresponding eigenstate distribution shows that the defect states either vanish or move to one of the end nodes along with all bulk and edge states in the presence of non-Hermiticity. Common parameters used: $t_1=1.2$, $t_2=4$ and $\gamma=0$.  }
  \label{fig6}
\end{figure}  
Now, we investigate the interplay between non-Hermiticity and topological defect states by studying a finite SSH chain  with a defective node  (indicated by the orange circle) that disrupts the alternating series of $t_1$ and $(t_2 \pm \delta)$ couplings between neighboring nodes (refer to Fig. \ref{fig6}a). For a start, we assume there is no gain/loss term. We plot the corresponding eigenvalue and eigenstate distributions both with and without non-Hermiticity in Fig. \ref{fig6}b-e.

In the absence of non-Hermiticity (i.e., $\delta=0, \gamma=0$), we observe zero-energy edge and defect states, as shown in Fig. \ref{fig6}b. These defect states are localized at the defective node, and there are two edge states in the eigenstate distribution, which are respectively localized at the two edges as shown  in Fig. \ref{fig6}d. We then insert a finite non-Hermiticity into the system by setting the non-reciprocity factor $\delta$ to a finite value. Interestingly, although the eigenenergy spectrum remains largely unchanged (compare Fig. \ref{fig6}b with Fig. \ref{fig6}c), the defect states either vanish or move to one of the edges along with all of the bulk states which comprise of the NHSE localization (see Fig. \ref{fig6}e). Additionally, one of the topological edge states also shift to the other edge, joining the other edge state, as well as defect and bulk states localized there. Consequently, one can control the emergence and disappearance of the topological defect states merely by turning on or off the non-Hermiticity of the system.

Note that, in the homogenous state without defect, the left and right edge states are degenerate, and they will be dragged to the preferred side of NHSE. However, if the degeneracy of the left and right edge states are broken, the edge states will not be dragged by the NHSE, contrary to conventional expectations. As discussed in details in the next section, the breaking of degeneracy may occur in a heterojunction system with defect states and gain/loss potentials corresponding to different PT symmetries (symmetric, antisymmetric and broken). Additionally, we also note that the NHSE leads to the localization of all states towards the preferred edge only if there is matching non-Hermitian coupling asymmetry parameter across the defect states. Conversely, if the decay length of the non-Hermitian bulk states on either side of the defect states is unequal, the bulk states experiencing NHSE in the two segments will not converge to one edge by crossing the defect states. 

\section{Effect of gain/loss on the defect states}

\begin{figure*}[ht!]
  \centering
    \includegraphics[width=0.7\textwidth]{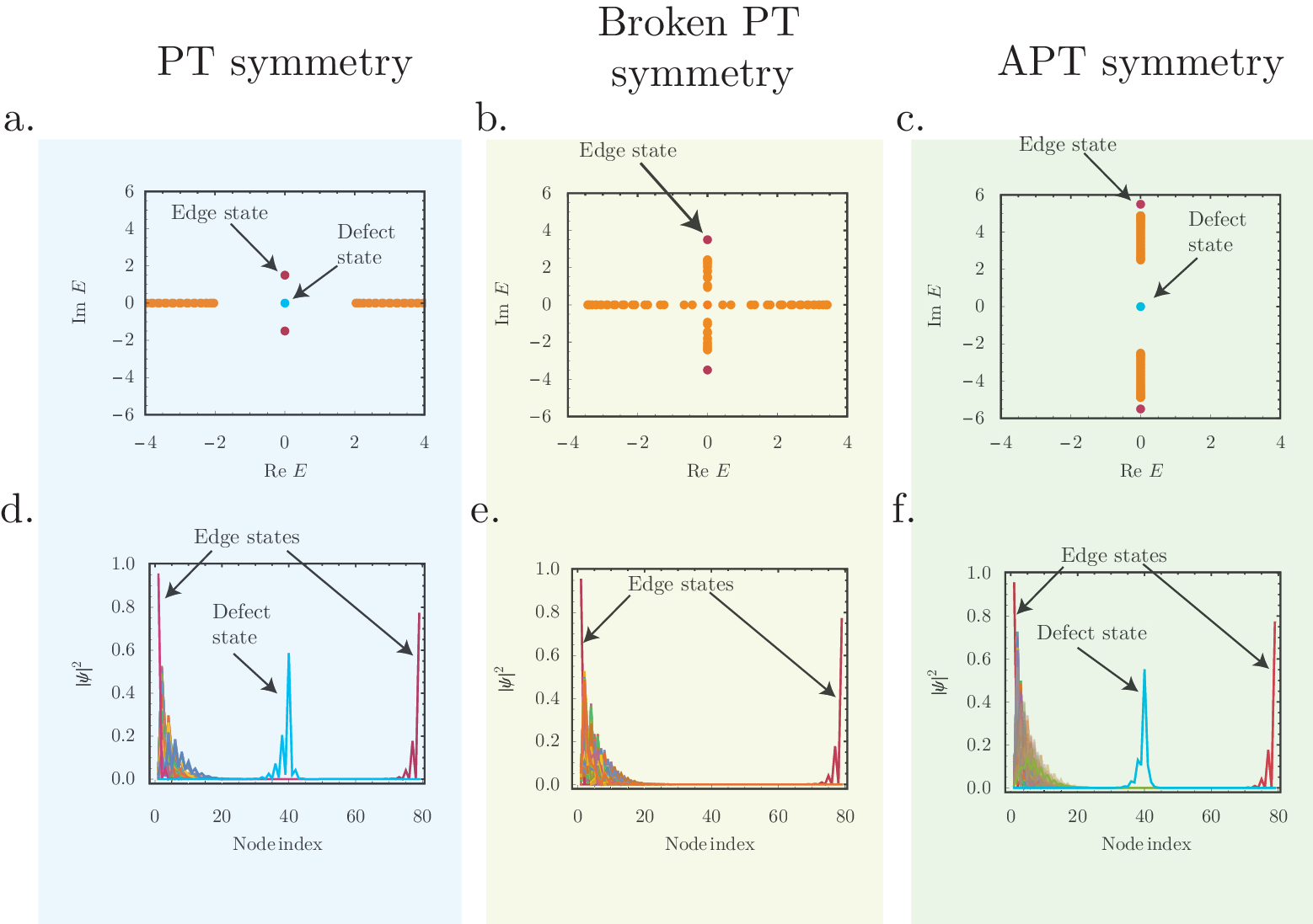}
  \caption{Evolution of defect states at different symmetry cases realized by varying the balanced gain/loss $\gamma$. The complex energy spectra and eigenstate spatial distributions for the PT-symmetric ($\gamma=1.5$), BPT-symmetric ($\gamma=3.5$), and APT-symmetric ($\gamma=5.5$) cases are shown in (a)-(c), respectively. The corresponding eigenstates distributions are shown in (d)-(f). The defect states are observed only in the PT- and APT-symmetric cases (d) and (f) while they are absent in the broken PT-symmetric case (e). The defect states appear at $E=0$ while the edge states are localized at $E=\pm i\gamma$. Interestingly, the presence of gain/loss does not drag the defect states into the extensively localized bulk states. Common parameters used: $t_1=1.2$, $t_2=4$, and $\delta= -1.5$.  }
  \label{fig7}
\end{figure*}

We now investigate the effect  of gain/loss terms on the characteristics of the defect states \cite{munoz2018topological,zurita2023fast}. To do so, we add a balanced gain/loss term to the $\mathrm{A}/\mathrm{B}$ nodes respectively, while leaving the defect node passive, as shown in Fig. \ref{fig6}a. As before, by varying the $\gamma$ factor, we can  realize all three PT symmetries, i.e., the PT-symmetric, BPT-symmetric, and APT-symmetric cases, the eigenspectra of which are depicted  in Figs. \ref{fig7}a to \ref{fig7}c, respectively. Interestingly, in the presence of a finite $\gamma$, defect states survive only in the PT- and APT-symmetric cases (see Fig. \ref{fig7}d,f), but are absent in the broken PT-symmetric case (see Fig. \ref{fig7}e). Energy-wise, the defect states are located at $E=0$, while the edge states are located at $E=\pm i\gamma$. In all cases, we simultaneously observe both the  NHSE and topological edge state localization.

\begin{figure*}[ht!]
  \centering
    \includegraphics[width=0.7\textwidth]{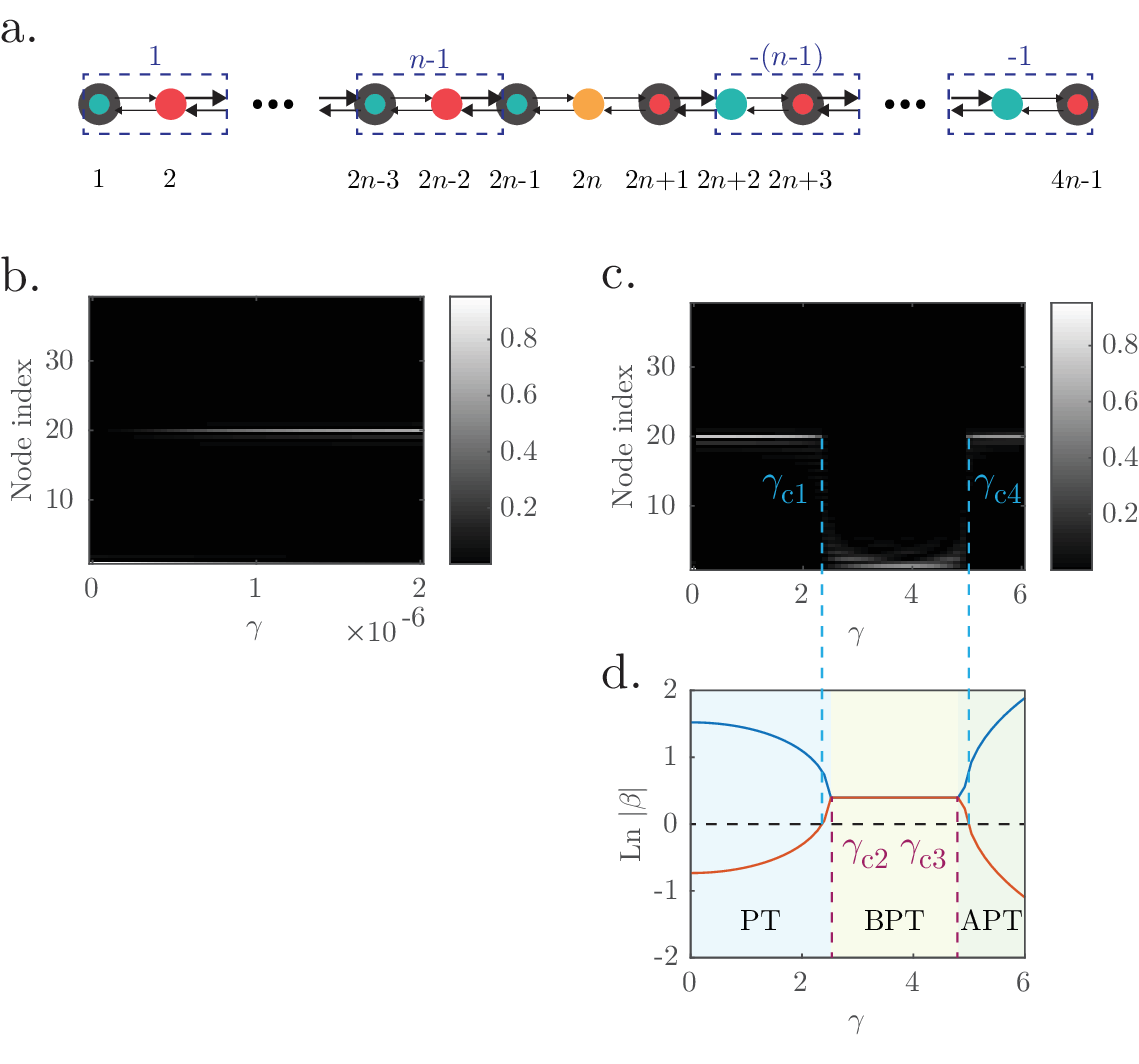}
  \caption{Evolution of zero-energy states with balanced gain/loss term $\gamma$. (a) shows a schematic representation and labeling scheme of a finite system containing $(n-1)$ complete unit cells on the left and right of the nodes on either side of the defect (orange). Each complete unit cell is denoted by a dashed box. The complete unit cells are numbered from 1 to $(n-1)$ for the cells to the left of the defect, and from $-(n-1)$ to -1 on the right of the defect (blue labels above chain). The nodes of the coupled system are labeled from 1 to $4n-1$ (black numbers below chain). The nodes with the thick borders have finite probability densities at $\gamma=0$.  (b) The spatial probability density profiles of the zero-energy states for a $n=10$ system with the parameter set of Fig. \ref{fig7}, i.e. $t_1=1.2$, $t_2=4$, and $\delta =-1.5$ as a function of $\gamma$ for small $\gamma$ values. (c) The corresponding probability density profiles across a wider $\gamma$ range, and (d) the values of $\mathrm{Ln} |\beta_{>}|$ (blue curve) and $\mathrm{Ln}|\beta_{<}|$ (red curve) as functions of $\gamma$. The vertical dotted lines denote the critical values of $ \gamma  = \gamma_{c1}, \gamma_{c4}$ between which  $\mathrm{Ln} |\beta_{>}|$ and $\mathrm{Ln} |\beta_{<}|$ have the same signs and demarcate the switch-overs  in the localization direction between the right edge of the system and the defect site, and $\gamma=\gamma_{c2}, \gamma_{c3}$ between which $|\beta_{>}|=|\beta_{<}|$.    }
  \label{gFig5}
\end{figure*}  

To explain these observations in detail, we adopt the labeling scheme in Fig. \ref{gFig5}a for a finite open chain containing $n-1$ complete unit cells (i.e., a pair of A and B-sublattice nodes) on either side of the three nodes comprising the defect and its immediate neighbors. We number the \textit{unit cells} from 1 to $n-1$ for the unit cells to the left of the defect, and from $-(n-1)$ to -1 for the unit cells on the right of the defect (blue labels above the chain in Fig. \ref{gFig5}a). We concurrently number the \textit{individual nodes} in the chain from 1 to $4n-1$ (black labels below the chain in Fig. \ref{gFig5}a). 

Denoting the state of the entire chain as $|\psi\rangle$, we write the wavefunction of the $m$th complete unit cell on the left of the defect as $\langle m|\psi_{\mathrm{L}}\rangle$ where $m$ ranges from 1 to $n$, that of the $m$th complete unit cell on the right of the defect as $\langle m|\psi_{\mathrm{R}}\rangle$ where $m=-(n-1)$ to -1, and similarly that of the $x$th \textit{individual node} as $\langle x|\psi\rangle$. Note then that in this labeling scheme, $\langle x=2(n-1)|\psi\rangle = \langle \mathrm{B}| \langle m=(n-1)|\psi_{\mathrm{L}}\rangle$, and $\langle x=2(n+1)|\psi\rangle = \langle \mathrm{A}| \langle m = -(n-1)|\psi_{\mathrm{R}}\rangle$, as indicated in Fig. \ref{gFig5}a.  

At $\gamma=0$, Eq. \eqref{eqH0} for a periodic system can be written in the Bloch matrix form as 
\begin{equation}
	H_0(\beta) = \begin{pmatrix} 0 & t_1 + (t_2+\delta)/\beta \\   t_1 + (t_2-\delta)\beta & 0 \end{pmatrix}. \label{H0b} 
\end{equation} 
 We observe from our numerical results that when defect states do exist in the defective chain, these defect states are pinned at the eigenenergy of $E=0$ \cite{marques2022generalized}. We therefore focus on the zero-energy eigenstates. Eq. \ref{H0b} admits two zero-energy eigenvectors, one of which is $(0, 1)^\mathrm{T}$  at $\beta^{(0)}_< = -(t_2+\delta)/t_1$ and the other one of which is $(1, 0)^\mathrm{T}$ at $\beta^{(0)}_> = -t_1/(t_2-\delta)$ where the superscript $(0)$ refers to the $\gamma=0$ case here.  The wavefunctions at the $m$th unit cell to the left and right of the defect can thus be written as
\begin{align}
	\langle m|\psi^{(0)}_{\mathrm{L}}\rangle &= (\beta^{(0)}_<)^m \begin{pmatrix} 1 \\ 0 \end{pmatrix} \label{psi0L} \\
\langle m|\psi^{(0)}_{\mathrm{R}}\rangle &=  c^{(0)}_{>,\mathrm{R}} (\beta^{(0)}_>)^m \begin{pmatrix} 0 \\ 1 \end{pmatrix} \label{psi0R}
\end{align}
where $c^{(0)}_{>,\mathrm{R}}$ is an unknown weight to be solved for. 

Notice that $|\psi^{(0)}_{\mathrm{L}}\rangle$ ($|\psi^{(0)}_{\mathrm{R}}\rangle$) satisfies the boundary condition that the wavefunction must vanish at the node immediately to the left (right) of the leftmost (rightmost) node of the chain, i.e., $\langle \mathrm{B}|\langle m=0|\psi^{(0)}_{\mathrm{L}}\rangle = 0$ ($\langle \mathrm{A}|\langle m=0|\psi^{(0)}_{\mathrm{R}}\rangle = 0$). To obtain the complete wavefunction of the chain, we apply the condition that the wavefunction must satisfy the Schroedinger equation $\langle x|H_0|\psi^{(0)}\rangle = 0$ at $x=2n-1$, $2n$, and $2n+1$ which leads to the following equations:

\begin{align} 
	(t_2+\delta)\langle x = 2n-2|\psi^{(0)}\rangle + t_1 \langle x = 2n|\psi^{(0)}\rangle & = 0 \label{SEp2nm1} \\  
	t_1 (\langle x=2n-1|\psi^{(0)}\rangle + \langle x=2n+1|\psi^{(0)}\rangle) &= 0 \\
	t_1 \langle x=2n|\psi^{(0)}\rangle + (t_2-\delta) \langle x=2n+2|\psi^{(0)}\rangle & = 0 \label{SEp2np1}
\end{align} 

Recalling that $\langle x = 2n-2|\psi^{(0)}\rangle = \langle \mathrm{B}|\langle m=n-1|\psi^{(0)}_{\mathrm{L}}\rangle$ and $\langle x = 2n+2|\psi^{(0)}\rangle = \langle \mathrm{A}|\langle m=-(n-1)|\psi^{(0)}_{\mathrm{R}}\rangle$ (refer to Fig. \ref{gFig5}a) and using the definitions of $|\psi^{(0)}_{\mathrm{L/R}}\rangle$ in Eqs. \eqref{psi0L} and \eqref{psi0R}, the solution of Eqs. \ref{SEp2nm1} to \ref{SEp2np1} yields
\begin{align}
	\langle m|\psi^{(0)}_{\mathrm{L}}\rangle &= \left( -\frac{t_1}{t_2-\delta} \right)^m \begin{pmatrix} 1 \\ 0 \end{pmatrix}  \label{p0Lm} \\
	\langle x=2n-1|\psi^{(0)}\rangle &= \left( -\frac{t_1}{t_2-\delta} \right)^n \\
	\langle x=2n|\psi^{(0)}\rangle &= 0 \\
	\langle x=2n+1|\psi^{(0)}\rangle &= -\left( -\frac{t_1}{t_2-\delta} \right)^n \\
	\langle m|\psi^{(0)}_{\mathrm{R}}\rangle &= -\left( \frac{\delta-t_2}{\delta + t_2} \right)^n\left( -\frac{t_2-\delta}{t_1} \right)^m \begin{pmatrix} 0 \\ 1 \end{pmatrix} \label{p0Rm}. 
\end{align}

These expressions imply that at $\gamma=0$, only alternating nodes starting from the leftmost node in the chain have finite  probability densities, as denoted by the thick black boundaries around these nodes in Fig. \ref{gFig5}a. The defect node here serves as a transition point between the unit cells to its left, in which only the A sublattice nodes have finite probability densities, and the unit cells to its right, in which only the B sublattice nodes have finite probability densities. In particular, from Eqs. \eqref{p0Lm} and \eqref{p0Rm}, the wavefunctions at the leftmost node $\langle x=1|\psi^{(0)}\rangle = \langle A| \langle m=1|\psi^{(0)}\rangle$ and rightmost node  $\langle x =4n-1|\psi^{(0)}\rangle = \langle B| \langle m=-1|\psi^{(0)}\rangle$ are given by
\begin{align}
	\langle x=1|\psi^{(0)}\rangle &= -\frac{t_1}{t_2-\delta} \\
	\langle x=4n-1|\psi^{(0)}\rangle  &= \left(\frac{ t_2 + \delta }{t_2-\delta}\right)^n \frac{t_1}{t_2+\delta}
\end{align}
For $|\delta| < |t_2|$ and $\delta$ having an opposite sign to $t_2$, as is the case here, the pre-factor of $((t_2+\delta)/(t_2-\delta))^n$ in $\langle x=4n-1|\psi^{(0)}\rangle$ implies that $|\langle x=1|\psi^{(0)}\rangle| > |\langle x=4n-1|\psi^{(0)}\rangle|$ leading to eigenstate localization on the left edge of the chain, as shown in Figs. \ref{fig6}e and \ref{gFig5}b. 

When $\gamma$ has a finite value, the zero-energy eigenvectors of $H(\beta)$ become $(1, \alpha_<)^{\mathrm{T}}$ at $\beta = \beta_<$ and $(\alpha_>, 1)^{\mathrm{T}}$ at $\beta = \beta_>$ where $|\beta_>| > |\beta_<|$ and, for $\delta^2+\gamma^2 -t_1^2 - t_2^2 < 0$, 
\begin{widetext}
\begin{align}
	\beta_{<, >} &= \frac{ \gamma^2 + \delta^2 - t_1^2 - t_2^2 \pm \sqrt{ ((t_1-\gamma)^2 + \delta^2 - t_2^2)((t_1+\gamma)^2 + \delta^2 - t_2^2) } }{2 t_1(t_2-\delta)} \\
	\alpha_{<, >} &= \pm \frac{ 2i t_1 \gamma }{t_2^2 - t_1^2 - \gamma^2 - \delta^2 + \sqrt{ ((t_1-\gamma)^2 + \delta^2 - t_2^2)((t_1+\gamma)^2 + \delta^2 - t_2^2) }}.
\end{align}
\end{widetext}
Notice that the $\alpha_{<,>}$s are proportional to $\gamma$ and vanish when $\gamma=0$. 
 
To satisfy the boundary conditions that the wavefunction must vanish at the node immediately to the left of the leftmost node of the chain (i.e., $\langle \mathrm{B}|\langle m=0|\psi_{\mathrm{L}}\rangle = 0$) and at the node immediately to the right of the rightmost node of the chain (i.e., $\langle \mathrm{A}|\langle m=0|\psi_{\mathrm{R}}\rangle = 0$), $\langle m|\psi_{\mathrm{L/R}}\rangle$ assume the forms of 
\begin{align}
	\langle m|\psi_{\mathrm{L}}\rangle = \beta_<^m \begin{pmatrix} 1 \\ \alpha_< \end{pmatrix} - \alpha_< \beta_>^m \begin{pmatrix} \alpha_> \\ 1 \end{pmatrix}, \\
	\langle m|\psi_{\mathrm{R}}\rangle = c_{>,\mathrm{R}} \left( \beta_>^m \begin{pmatrix} \alpha_> \\ 1 \end{pmatrix} - \alpha_> \beta_<^m \begin{pmatrix} 1 \\ \alpha_< \end{pmatrix} \right).
\end{align}

It can then be seen that the wavefunctions two nodes to the left ($\langle x=2(n-1)|\psi\rangle = \langle \mathrm{B}| \langle m = (n-1)|\psi_{\mathrm{L}}\rangle$) of the defect and two nodes to the right of the defect ($\langle x=2(n+1)|\psi\rangle = \langle \mathrm{A}| \langle m = -(n-1)|\psi_{\mathrm{R}}\rangle$) are respectively given by 
\begin{align}
	\langle x = (2n-1)|\psi\rangle &\approx \alpha_< \beta_>^{(n-1)}  \label{ps2nm1} \\
	\langle x = (2n+1)|\psi\rangle &\approx c_{>, \mathrm{R}} \beta_<^{-(n-1)} \label{ps2np1}
\end{align}
where we have neglected terms proportional to $\beta_<^{n-1}$ and $\beta_>^{-(n-1)}$  in Eqs. \eqref{ps2nm1} and \eqref{ps2np1}, respectively, because these terms are much smaller in comparison. 
These expressions imply that when $|\beta_>|> 1$ and $|\beta_<| < 1$, as is the case here, the wavefunction amplitudes around the defect become exponentially larger than the ones at the edges ( $|\beta
_< - \alpha_>\alpha_<\beta_>|$ and $|c_{>,\mathrm{R}} \beta_> - \alpha_>\alpha_<\beta_<|$, respectively) by the factors of approximately $|\beta_>|^{n-2}$ and $|\beta_<|^{-(n-2)}$, respectively. This results in the localization of the wavefunction around the defect with the increase of $\gamma$ (since both $\alpha_>$ and $\alpha_<$ are proportional to $ \gamma$), as shown by the shift of the white line denoting high probability density from node index $x = 1$ to $x = 20$, which corresponds to the node index of the defect site when $\gamma$ exceeds a value of approximately $0.5\times 10^{-6}$ (see Fig. \ref{gFig5}b). This can be shown more explicitly by solving  the set of equations for the Schroedinger equation at nodes $2n-1$ to $2n+1$  analogous to Eqs. \eqref{SEp2nm1} to \eqref{SEp2np1} which yield the expressions for the wavefunctions at the leftmost ($\langle x=1|\psi\rangle$), defect ($\langle x=2n|\psi\rangle$), and rightmost ($\langle x=4n-1|\psi\rangle$) nodes  as  

\begin{widetext}
\begin{align}
	|\langle x=1|\psi\rangle| =& \left |\beta_< - \alpha_>\alpha_< \beta_> \right| \label{eq38} \\
	|\langle x = 2n|\psi\rangle| =& \left| \frac{\alpha_< }{\beta_>\alpha_>\beta_<\alpha_<t_1(t_2-\delta)}\left( \beta_>^{n} \beta_< \alpha_> (t_2^2 + i \alpha_< t_1-\gamma^2 - \delta^2) + \beta_> \beta_<^n ( \alpha_> (\gamma^2+\delta^2-t_2^2) - i t_1\gamma) \right) \right| \label{eq39}\\
	|\langle x=4n-1|\psi\rangle| =& \left|\frac{\alpha_<}{\alpha_>}\frac{t_2+\delta}{t_2-\delta} \frac{1}{(\beta_>\beta_<)^{n+1}} (\beta_> - \alpha_>\alpha_< \beta_<) \right|. \label{eq40}
\end{align} 
\end{widetext}

Notice that at large $n$, the wavefunction at the defect (see Eq. \ref{eq39}) becomes exponentially larger than the wavefunctions at the left and right ends of the system (Eq. \ref{eq38} and Eq. \ref{eq40}) because of the factor of $\beta_>^n$ in the former. This results in the localization of the wavefunction in the vicinity of the defect site for much of the PT regime, as shown in \ref{fig7}a and Fig. \ref{gFig5}c.

The above arguments for the localization of the wavefunction around the defect are contingent on $|\beta_>| > 1$ and $|\beta_<| < 1$ and break down when these conditions no longer hold. As shown in  Fig. \ref{gFig5}d, the values of $|\beta_>|$ and $|\beta_<|$ converge as $\gamma$ is increased from 0. At the critical value of $|\gamma| = \sqrt{ (|t_2|-|t_1|)^2 - \delta^2} \equiv \gamma_{c1}$, both $|\beta_>|$ and $|\beta_<|$ become larger than 1. When this occurs, the wavefunction is no longer localized around the defect because localization at the defect requires the wavefunction to the left of the defect to be localized to the right, i.e., $|\beta_>| > 1$, and the wavefunction to the right of the defect to be localized to the left, i.e. $|\beta_<|< 1$. Instead, when both $|\beta_>|$ and $|\beta_<|$ are larger (smaller) than 1, we have the situation where the wavefunction is localized at the right (left) edge of the entire chain (see Fig. \ref{gFig5}c). 

As $\gamma$ is increased further beyond $|\gamma| = |\gamma_{c2}|= \sqrt{t_2^2-\delta^2}-|t_1|$, $|\beta_>|$ becomes equal to $|\beta_<|$. This corresponds to the BPT phase at which $E=0$ falls within the generalized Brillouin zone, which is defined as the loci of energies at which $|\beta_<|=|\beta_>|$. Since  $|\beta_<|=|\beta_>|$, this implies that both are simultaneously either larger or smaller than 1, and hence the zero-energy eigenstates are localized at either the right or left edge of the chain but not around the defect in the BPT regime, as shown in\ref{fig7}b and  Fig. \ref{gFig5}b. As $\gamma$ is increased further, $|\beta_<|$ and $|\beta_>|$ diverge when $\gamma$ is increased beyond $|\gamma|=|\gamma_{c3}| = \sqrt{t_2^2-\delta^2}+|t_1|$ and the system enters the APT regime. When $\gamma$ is increased further beyond $|\gamma| =\sqrt{ (|t_1|+|t_2|)^2 - \delta^2} \equiv \gamma_{c4}$, $|\beta_>|$ becomes larger than 1 while $|\beta_<|$ smaller than 1, and the wavefunction becomes localized around the defect again (Fig. \ref{fig7}f).   Thus, whether the wavefunction localization occurs at the defect site or at the system boundaries depend intimately on the eigenspectrum and its location in the GBZ, and the corresponding PT-symmetry of the system.

\section{Proposal for the Experimental Realization of Defect States in a Topolectrical Circuit}

\begin{figure*}[ht!]
	\centering
	\includegraphics[width=0.85\textwidth]{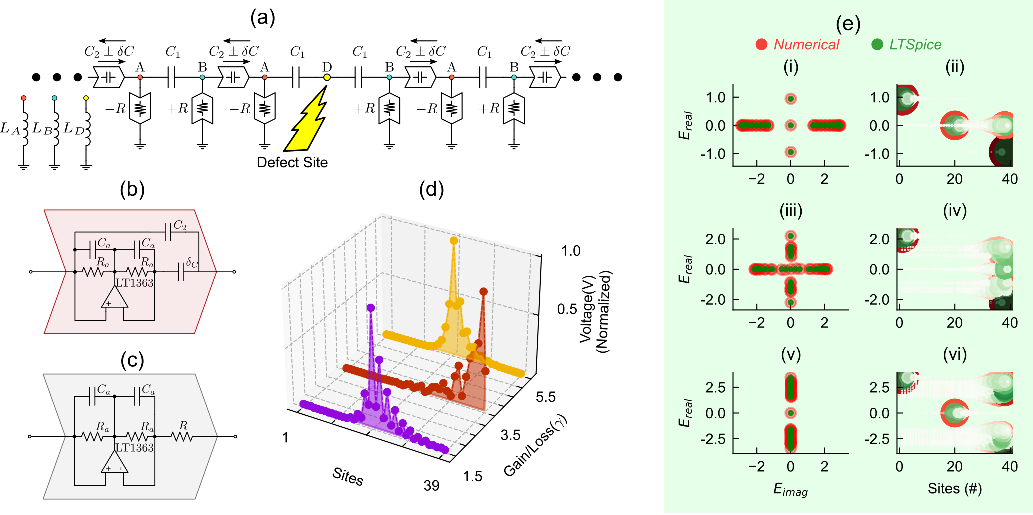}
	\caption{
		(a) Schematic of the electrical circuit realization of defect states for the proposed model. There are a total of $39$ nodes. Nodes within the unit-cell are connected by reciprocal capacitor $C_1$, while nodes from one unit cell to the neighboring unit cell are connected by non-reciprocal capacitor $C_2\pm\delta C$ realized by INIC. All $A$ nodes are grounded by inductor $L_A$ in parallel with $-R$ realized by INIC, and all $B$ nodes are grounded by inductor $L_B$ in parallel with $+R$ realized by INIC. Node $D$ is the defect node, grounded only through an inductor $L_D$. $C_1=1.2$ nF, $C_2=4$ nF, $\delta C=1.5$ nF were used for the intra- and inter-cell capacitive couplings, respectively. 
		(b-c) INIC module to realize non-reciprocal capacitance and negative resistance. (d) Normalized node voltage profiles for three different values of $\gamma$: 1.5 (\text{PT phase}), 3.5 (\text{BPT phase}), and 5.5 (\text{APT phase}). (e) Eigen-admittance spectra and admittance eigenstate distribution for (i)-(ii) PT symmetric, (iii)-(iv) BPT symmetric, (v)-(vi) APT symmetric cases (note that the admittance energy scale is normalized to units of $\mu \Omega^{-1}$). Red circles denote numerical results, while green circles denote results from LTSpice simulations. In (ii), (iv), (vi), the size of the circle represents the magnitude of the eigenstate at that particular site with real eigenenergy along the $y$-axis.
	}
	\label{figCkt}
\end{figure*}

We designed an electrical circuit to realize the proposed model, and verify the interplay between non-Hermiticity, edge states, and topological defect states. Fig. \ref{figCkt}(a) shows the circuit schematic with a total of $39$ nodes, including a defect site node $D$ in the middle. All other nodes are denoted as either $A$ or $B$. Intra-cell coupling between nodes $A$ and $B$ is achieved with reciprocal capacitance $C_1$, while inter-cell coupling is achieved with non-reciprocal capacitance $C_2\pm\delta C$. The non-reciprocal capacitance is realized using impedance converters at current inversion (INIC)\cite{rafi2021topological,rafi2022critical,rafi2022unconventional,rafi2022interfacial}. 

To drive the circuit at a frequency of $f_r=100$ kHz, nodes $A$, $B$, and $D$ are grounded through inductors $L_A$, $L_B$, and $L_D$ respectively, where $L_A=1/(\omega^2_r (C_1+C_2+\delta_C))$, $L_B=1/(\omega^2_r (C_1+C_2-\delta_C))$, and $L_D=1/(2\omega^2_r C_1)$. To ensure consistency with our numerical results, we used $C_1=1.2$ nF, $C_2=4$ nF, $\delta C=1.5$ nF, $L_A=0.685$ mH, $L_B=0.378$ mH, and $L_D=1.055$ mH. Additionally, to incorporate the on-site gain and loss term $\gamma$, we utilized positive and negative resistances ($\pm R$) using INIC. To operate the circuit in PT, BPT, and APT regimes at the resonant frequency of 100 kHz, we used $R=1061.03 \Omega$, $454.73 \Omega$, and $289.37 \Omega$ respectively. Figs. \ref{figCkt}(b) and (c) show the circuit schematic for non-reciprocal capacitance $C_2\pm\delta C$ and $\pm R$ using INIC. We chose $R_a=100 \Omega$ and $C_a=10$ pF after analyzing the stability of the Operational Amplifier operating at $f_r=100$ kHz.

Fig. \ref{figCkt}(d) shows the normalized node voltage profile when the circuit nodes were excited with current sources. For the PT symmetric case ($R=1061.03 \Omega$, $\gamma=1.5$), high voltage accumulation at the defect nodes confirms the theoretical predictions in \ref{fig7}(d). When the circuit operates in the Broken PT symmetric regime ($R=454.73 \Omega$, $\gamma=3.5$), the defect node voltages disappear, appearing only at the right edges due to the presence of NHSE. However, under the APT symmetry ($R=289.37 \Omega$, $\gamma=5.5$), the defect voltage accumulation reappears despite the presence of NHSE.

Fig. \ref{figCkt}(e) shows the  admittance eigenvalue spectrum and eigenstates reconstructed from the LTSpice simulations, compared with numerically calculated results. Red circles represent the numerical results, and green circles denote results from LTSpice simulations. By applying Kirchhoff's current law, the circuit system response can be described by \(\mathbf{V} = \mathbf{G} \mathbf{I} = \mathbf{J}^{-1} \mathbf{I}\), where \(\mathbf{J}\) is the admittance matrix or circuit Laplacian, \(\mathbf{G} = \mathbf{J}^{-1}\) is the circuit Green's function, and the vector components of \(\mathbf{I}\) and \(\mathbf{V}\) correspond to the input currents and voltages at the nodes. For an alternating current with angular frequency \(\omega = 2 \pi f\), the circuit Laplacian under periodic boundary conditions can be related to the lattice Hamiltonian $H(k)$ by the relation, $\mathbf{J}(\omega) = \frac{\mathbf{I}(\omega)}{\mathbf{V}(\omega)} = i \omega H (k)$. To reconstruct the circuit Laplacian, we injected current to each node separately and measured all node voltages. These node voltages represent the first column of the circuit Green's function, $G=J^{-1}$ with $J$ being the circuit Laplacian. 

For example, Fig. \ref{figCkt}(e) (i)-(ii) shows the comparison for the circuit operating in the PT symmetric regime. Fig. \ref{figCkt}(e)(i) shows that the theoretical admittance eigenspectrum (red) matches well with the energy values obtained from the LTSpice simulations (green). The eigenstates from the circuit Laplacian also match well with the theoretical results. In Fig. \ref{figCkt}(e)(ii), higher accumulation at the defect node (Node $20$) with zero admittance energy is observed, while other states with zero admittance accumulate along the right edge. Additionally, there is edge state accumulation with admittance value of $\pm 1$ (normalized unit of $\mu \Omega ^{-1}$). Fig. \ref{figCkt}(e)(iv) shows the results for the Broken PT symmetric case, where eigenstate accumulation at the defect node is no longer observed. However, in the APT case, the defect state reappears. These results confirm the practical implementation of the proposed defect model via a topolectrical circuit. As electrical circuits are versatile in design and operation, this implementation can be useful for practical applications.

\section{Conclusion}
In conclusion, we have explored the interplay between NHSE and gain/loss terms with the topological edge and defect states in a finite SSH chain. As is well-known, the SSH chain hosts topological edge states localized at both ends of the chain. In the presence of conventional NHSE induced by non-reciprocity, both the topological edge states as well as bulk states are dragged towards a single preferred edge. However, with the introduction of staggered gain and loss terms to the sublattice sites, the topological edge states recover their original localization at both edges, while the bulk states retain their localization at the preferred edge. We also introduced defect sites in the chain and analyzed the properties of the resulting defect states under the influence of the NHSE and the underlying PT symmetry of the system. Interestingly, we found that the NHSE as well as the PT symmetry can modulate the localization of the defect state to occur either at the defect site or at the edges of the SSH chain, or to suppress the localization of the defect state completely. This is in contrast to the Stegmaier \textit{et al.} (Ref. \onlinecite{stegmaier2021topological}), which focused primarily on the activation / deactivation of defect states through gain / loss engineering. We also provided the analytical basis of this phenomenon of NHSE-controlled localization of defect states by evaluating the spatial profile of the eigenstates. Moreover, we proposed an experimental realization of our model using a TE circuit constructed from basic components such as capacitors, inductors, resistors and operational amplifiers. This circuit design validates our theoretical predictions by demonstrating the presence and control of defect states through its electrical characteristics, including voltage and admittance responses, modulated by non-Hermitian parameters. The ability to switch topological defect states and adjust their localization has significant implications for defect engineering in non-Hermitian systems. Our study provides new insights into the interplay between non-hermiticity and topological defect states,  for the design and engineering of non-Hermitian systems with controllable and switchable defect states.

\subsection*{Acknowledgments}

This work is supported by the Ministry of Education (MOE) of Singapore Tier-II Grant MOE-T2EP50121-0014 (NUS Grant No. A-8000086-01-00), and MOE Tier-I FRC Grant (NUS Grant No. A-8000195-01-00).


%

\end{document}